\documentclass{article}

\usepackage{latexsym,amsfonts,amssymb,amsmath,amsthm}
\usepackage{graphics}
\usepackage[shortlabels]{enumitem}

\newcommand\al{\alpha}
\newcommand\ee{\mathbf e}
\newcommand\Ss{\mathbf S}

\newcommand\x{\mathbf x}
\newcommand\qq{\mathbf q}

\newcount\graphics\graphics=0  
\let\FT\mathrm

\begin{document}

\centerline{\Large{\bf On the origin of irreversibility}}
\medskip

\centerline{T. Matolcsi\footnote{Department of Applied Analysis and Computational Mathematics,
E\"otv\"os Lor\'and University, Budapest, Hungary}}

\begin{abstract}

The contradiction of micro-reversibility and macro-irreversibility is an old problem in statistical
mechanics. This article argues that irreversibility is present even at the micro level because of 
the mechanical-electromagnetic interaction, so there may actually be no contradiction.

\end{abstract}

\bigskip

\section{Introduction}

Statistical mechanics is an important and effective branch of modern physics which explains 
many thermodynamical relations (a comprehensive summary 
of a wide range of different results can be found e.g. in \cite{gasp}). 
It contains, however, a fundamental problem since its foundations: the contradiction between 
microscopic reversibility and macroscopic irreversibility.

Recent books state this problem very clearly. .
\medskip

``How does the irreversibility of macroscopic phenomena arise? 
It certainly does not come from the fundamental laws of physics, because these laws are all reversible.''
``How can we obtain an irreversible process from the combined effects of very many reversible processes? 
This is a vitally important question.''
(\cite{fitz}, p.37)

``How can a system obeying time-reversal-invariant laws of motion
show irreversible behavior? ''(\cite{swen}, p.5)
\medskip

It is customary to give an explanation for irreversibility by probabilistic considerations.

Maxwell was the first to state that irreversibility is stochastic, the second law is of the nature of
strong probability, not an absolute certainty. Then further researchers have accepted this conception,
although sometimes with reluctance, e.g. ``...though this law is not `fundamental' (since I essentially derived it by looking
at probabilities), I'll treat it as fundamental from now on.'' (\cite{schr}, p.74) 

Further, it is stated  (\cite{viol1}, \cite{viol2},
\cite{viol3},\cite{alten}) that Axiom of Determinism is to be replaced by Axiom of 
Causality: "The future state of the system depends solely on the probabilities of events in the past".
Then a Fluctuation Theorem is derived from the Axiom of Causality, and
Second Law is obtained as a special case of Fluctuation Theorem.

However, we can be sure:
\medskip

\centerline{\parbox{103mm}{\it Probabilities that arise from our lack of knowledge
cannot explain a physical fact (irreversibility), which is independent of us;}} 
\medskip

on the contrary:
\medskip

\centerline{\parbox{95mm}{\it A physical fact (irreversibility), which is independent of us, can explain probabilities 
that arise from our lack of knowledge.}} 

\section{Action and interaction}

If there is a contradiction in a theory, we must have messed something up. Macroscopic irreversibility 
is an everday experience, so we may suspect that microscopic reversibility is at fault.
Let us find out if this is the case.

1. Statistical mechanics starts with classical mechanics: accepts that the movements of the particles  
are described by the Hamiltonian equation on the phase space. To determine a movement, we should but cannot solve about 
``$10^{24}$ simultaneous differential equations. Even if we could, we would not want to. After all, we are 
not particularly interested in the motions of individual particles. What we really want is statistical 
information regarding the motions of all particles in the system.'' (\cite{fitz}, p. 31)

2. The Hamiltonian function is the sum of the kinetic energies and the potential energies. The potentials describe
the interaction of the particles and can be of electromagnetic and other types, such as of Lennard -- Jones type. 
The interaction potential of the $i$-th and $k$-th particle is of the form $V(\qq_i-\qq_k)$ where $\qq$-s are 
positions in the phase space. Such potentials describe instantaneous interaction at a distance.

3. Instantaneous interaction at a distance is a non-relativistic concept. Non-relativistic mechanics provides
a simplified and distorted picture of reality. Relativity theory revealed:
\medskip

\centerline{\parbox{90mm}{\it Instantaneous interaction at a distance does not exist; 
fields mediates interaction between particles, or, better to say, there are interactions
between particles and fields.}}
\medskip

4. Although it is customary to say interaction between 
electromagnetic field and matter, the usual formulas of electromagnetic phenomena 
(e.g. \cite{jackson}, \cite{hutter}, \cite{groot}) describe only actions and some of them are not well posed (\cite{matt}). 

There is no doubt that both classical mechanics and electrodynamics  in their known forms are 
{\bf theories of given actions}: 

-- \ the motion of charged particles in a {\bf given electromagnetic field} is described by the Newtonian
equation with the Lorentz force,

-- \ the radiated electromagnetic field due to a {\bf given motion of charged particles} is described
by the Maxwell equations
\smallskip

\noindent but

\smallskip

-- \  we do not have an equation to describe the mechanical-electromagnetic interaction i.e.  
{\bf processes consisting of the motion of particles and the propagation of electromagnetic field together},
This is reflected in the problem of radiation reaction force and the Lorentz--Dirac equation (\cite{mat}).  

After all, we can sate:
\medskip

\centerline{\parbox{95mm}{\it Instead of particle-particle interaction (Hamiltonian equation), 
particle-field interaction must be taken when treating microscopic processes in thermodynamics.}}
\medskip

The trouble is that at present we have no description of particle-field interactions.

5. One would think that quantum mechanics, the fundamentally probabilistic theory,
 is more suitable than classical mechanics. However, 

-- \ quantum statistical formulas treat only
transition probabilities between eigenstates of energy and a spin component (not arbitrary ones), 
and these transition probabilities 
follow the rules of classical probability theory,

-- \ quantum theory does not desribe particle-field interaction either.

\section{Reversibility and irreversibility}

Equations of mechanics (Newton, Hamilton, Schr\"odinger) as well as 
equations of electromagnetism (Maxwell), describing actions instead of interactions, 
suggest reversibility. But in reality, there is particle--field interaction for which, at present,
we have no equation. Therefore, the statements that ``... the fundamental laws of physics ... are 
all reversible'' and  ``... system obeying time-reversal-invariant laws of motion...'' are not justified. 
\medskip

\centerline{\parbox{100mm}{\it Irreversibility can be the consequence of particle-field interaction, based on 
the asymmetry of emission and absorption: \newline a particle emits radiation depending on its 
state but radiation for absorption arrives at it independently of its state.}}
\medskip

This supports the following {\bf hypothesis}: 
\medskip

\centerline{\framebox{\parbox{100mm}{\it Irreversibility is a fundamental law of Nature. Reversible laws of some
physical theories, such as mechanics, are only idealizations.}}}
\medskip

If this is true then irreversibility is present also at the microscopic level, so the contradiction 
of microscopic reversibility and macroscopic irreversibility disappears.
Then probability theory is relieved of the impossible task of resolving this contradiction and gains
a fair application. We have to use probabilities due to our lack of knowledge about the 
particle-field interaction.
\medskip

\section{Simple models for microscopic irreversibility}\label{model}

We consider the following simple one dimensional mathematical model for a body consisting of 
$N$ equal particles which, from a pure mechanical point of view, interact by equal elastic forces.
Then the Hamiltonian function is
$$\sum_{k=1}^N \frac{p_k^2}{2m} + (1-\delta_{k1})\frac{m\omega^2(q_k-q_{k-1})^2}{2} + 
(1-\delta_{kN})\frac{m\omega^2(q_k-q_{k+1})^2}{2}$$
and the motion of the particles would be reversible,

The particles in this body, however, do not move according to the Hamiltonian equation. 
Quantum mechanics gives the energy levels of harmonic oscillators and spectroscopic observations 
prove that particles can emit and absorb radiate electromagnetic waves whose frequencies correspond to the 
energy level differences. The particles in the body interchange energy by emitting and absorbing radiation 
continuously. The higher the energy level of a particle, the more energy can be radiated.

To illustrate such a process, we suppose that during given equal time periods each particle radiates a given fraction 
of its energy towards its neighbours and absorbs the radiation arriving at it. 
It is emphasized that this is a tentative assumption about the intaraction, not a statement. 
Then the $k$-th particle in a period 

\begin{itemize}[$\bullet$,nosep]
\item has average energy $e_k>0$,
\item radiates energy $\al e_k$ in both directions,
\item gains energy $\al e_{k-1} + \al e_{k+1}$ \end{itemize} 

\noindent where $k=1,\dots,N$ and $\al>0$ is `much smaller' than $1$, e.g. $0.01$.  
$e_{k-1}$ for $k=1$ and $e_{k+1}$ for $k=N$ will represent energies of the particles in diverse environments
that are in contact with the edge particles of the body.

Note that if $e_k>e_{k-1}$ then energy flows in a period from the $k$-th particle towards the $(k-1)$-th one.
                                                                                                 
The process of the particles -- a microscopic process of the body -- is the succession of such steps of radiation
and absorption.
\medskip

{\bf Body in a heat bath} The edge particles can exchange energy with an `infinitely large' environment whose 
particles have constant energy $e_a$. Then the first and last particle radiates energy $\al e_1$ and 
$\al e_N$, respectively, to the environment and they gain energy $\al e_a$, 

The state $\ee:=(e_1,\dots,e_N)$ in a period changes to  $S_a(\ee)$ in such a way that
 \begin{align}
\bigl(S_a(\ee)\bigr)_k &:= (1-2\al)e_k + \al(e_{k-1} + e_{k+1}), \quad k\neq1, N
 \\
\label{change1}\bigl(S_a(\ee)\bigr)_1 &:= (1-2\al)e_1 + \al(e_a + e_2), \\
\label{change2}\bigl(S_a(\ee)\bigr)_N &:= (1-2\al)e_N + \al(e_{N-1} + e_a).
 \end{align}

$$\ee_a:=\left(e_a,e_a,\dots,e_a\right)$$ 
is a fix point of $S_a$, so this state does not change by radiation; in other words, it is an equilibrium. Then

$$S_a(\ee)-\ee_a=S_a(\ee) - S_a\ee_a=:\Ss(\ee-\ee_a)$$
which means that $S_a$ is an affine map over the linear map $\Ss$ given by the matrix
\begin{equation}\label{irrev}\Ss:=\begin{pmatrix} 1-2\al & \al &  0 &  0 & \dots & 0 \\
\al & 1-2\al & \al & 0 & \dots &0 \\
0 & \al & 1-2\al & \al & \dots &0 \\
\vdots & \vdots & \vdots & \vdots & \vdots &\vdots \\
0 & 0 & \dots & \al & 1-2\al & \al \\ 
0 & 0 & \dots & 0 & \al & 1-2\al. \end{pmatrix}\end{equation}

This matrix (linear map) is a contraction (see the Appendix), so

$$\lim_{n\to\infty}\Ss^n(\ee-\ee_a)=0, \qquad \text{that is} \qquad 
\lim_{n\to\infty}S_a^n(\ee)=\ee_a:$$
`as time progresses' -- advancing with $S_\FT{a}$ step by step --, the equilibrium forced by the environment 
is established for any initial state. 
\medskip

\centerline{\parbox{72mm}{\it This model imitates the irreversible microscopic processes how
a body becomes homogeneous by taking on the temperature of the environment.}}
\medskip

{\bf Body between two heat baths} The edge particles can exchange energy 
with two `infinitely large' environments  whose particles have constant energies $e_a$ and $e_b$, respectively.
Then replacing $e_a$ in \eqref{change2} with $e_b$, we get $S_{ab}$ which is an affine map over the same linear 
map \eqref{irrev}. 

Supposing that $e_b>e_a$, with the notations
$$\Delta e_{ab}:=\frac{e_b - e_a}{N+1}$$
and
$$\ee_{ab}:=\Bigl(e_a+ \Delta e_{ab},\ e_a+ 2\Delta e_{ab},\ \dots,\ 
e_a+ N\Delta e_{ab}=e_\FT{b}-\Delta e_{\FT{a}\FT{b}}\Bigr)$$
we have that  

$$\lim_{n\to\infty}\Ss^n(\ee-\ee_{ab})=0, \qquad \text{that is} \qquad 
\lim_{n\to\infty}S_{ab}^n(\ee)=\ee_{ab}:$$
the states tend to the steady state $e_{ab}$
forced by the environments; the energy distribution of this state increases linearly.
\medskip

\centerline{\parbox{90mm}{\it This model imitates the microscopic processes of heat flow through a body from a 
warmer environment to a colder one; the processes, tending to a steady state, are irreversible.}}  
\medskip

{\bf Heat insulated body} The sum the particle energies is a constant $E>0$ and the possible states of the
particles are in 
$$Z_E:=\left\{\ee:=(e_1,\dots,e_N)\Bigm| \sum_{k=1}^N e_k=E\right\}$$
which is an affine space over the linear subspace
$$\mathbf Z_0:=\left\{\mathbf x \Bigm|\sum_{k=1}^Nx_k=0\right\}.$$

The energy radiated by the edge particles towards the heat insulation is reflected immediately to the particles.
Then replacing $e_a$ in \eqref{change1} and \eqref{change2} with zero, we get  $S_E$ which is an affine map
over the restriction  of the linear  map
$$\Ss_0:=\begin{pmatrix} 1-\al & \al &  0 &  0 & \dots & 0 \\
\al & 1-2\al & \al & 0 & \dots &0 \\
0 & \al & 1-2\al & \al & \dots &0 \\
\vdots & \vdots & \vdots & \vdots & \vdots &\vdots \\
0 & 0 & \dots & \al & 1-2\al & \al \\ 
0 & 0 & \dots & 0 & \al & 1-\al \end{pmatrix}$$
onto $\mathbf Z_0$; this restriction is a contraction (see the Appendix). Then with the notation

$$\ee_E:=\left(E/N,\dots,E/N\right)$$
we have that 

$$ \lim_{n\to\infty}\Ss_0^n(\ee-\ee_E)=0, \qquad \text{that is} \qquad 
\lim_{n\to\infty}S_E^n(\ee)=\ee_E:$$
the states tend to the equilibrium $\ee_E$ forced by the heat insulation.
\medskip

\centerline{\parbox{87mm}{\it This model imitates the irreversible microscopic processes 
how a heat insulated body becomes homogeneous.}}

\section{Discussion}

{\bf Micro-irreversibility}. Experiences and theoretical considerations show that particle-field interaction 
plays a significant role in the inner life of bodies. In this note we suggested that irreversibility is a 
consequence of this interaction. This hypothesis cannot be proven rigorously at present, because there is 
no available exact theoretical description of the particle-field interaction. Simple models imitating this 
interaction, however, support that the suggestion is true, so irreversibility can be a fundamental law 
even at the microscopic level. 

{\bf Fluctuations} If we knew the exact description of particle-field interaction then, instead of $\al e_k$, 
we would have time functions $t\mapsto\beta(t)e_k(t)$ whose mean value for a time period is $\alpha e_k$. 
Then it may happen that
$$e_k(t)>e_{k-1}(t) \quad \text{and} \quad \beta_k(t)e_k(t)<\beta_{k-1}(t)e_{k-1}(t)$$
for some $k$ and some  moment $t$ which could indicate -- without referring to probabilities -- 
that Clausius' formulation of the second law that ``heat cannot pass spontaneously 
from a colder to a warmer place'' may be violated locally and momentarily. 

{\bf Continuum thermodynamics} If irreversibility is indeed a fundamental law at the microscopic level, 
then the meaning and consequences of the second law can be researched and understood without any microscopic
background. From this point of view, it will be clear that in continuum thermodynamics 
the second law itself can result in evolution equations (\cite{vankov}), similar to the Euler--Lagrange equations obtained 
by variational principles for the non-dissipative case. 

\section{Appendix} 

Let us take the sum-norm on $N$-tuples: $\|\x\|:=\sum_{k=1}^N|x_k|$.
\medskip

1. $\Ss$ is a contraction. 

\begin{itemize}
\item $|(\Ss\x)_1|=|(1-2\al)x_1 +\al x_2|\le (1-2\al)|x_1| + \al|x_2|$, 

\item $|(\Ss\x)_N|=|(1-\al)x_N +\al x_{N-1}|\le (1-\al)|x_N| + \al|x_{N-1}|$,

\item $|(\Ss\x)_k|=|\al x_{k-1} + (1-2\al)x_k +\al x_{k+1}|\le \al |x_{k-1}| + (1-2\al)|x_k| +\al |x_{k+1}|$, 
for $k\neq 1,N$.
\end{itemize}

Then                                                                                

$$\sum_{k=1}^N|(\Ss\x)_k|\le \sum_{k=1}^N|x_k| -\alpha(|x_1| +|x_N|)<\|\x\|,$$
\medskip

2. The restriction of $\Ss_E$ to $\mathbf Z_0$ is a contraction.

\begin{itemize}
\item $|(\Ss_E\x)_1|=|(1-\al)x_1 +\al x_2|\le (1-2\al)|x_1| + \al|x_2|$ and equality holds if and only if
$x_1$ and $x_2$ have the same sign,

\item $|(\Ss_E\x)_N|=|\al x_{N-1} + (1-\al)x_N +|\le \al|x_{N-1}| + (1-\al)|x_N|$ and equality holds if and only if
$x_{N-1}$ and $x_N$ have the same sign,

\item $|(\Ss_E\x)_k|=|\al x_{k-1} + (1-2\al)x_k +\al x_{k+1}|\le \al |x_{k-1}| + (1-2\al)|x_k| +\al |x_{k+1}|$
and equality holds if and only if  $x_{k-1}$, $x_k$ and $x_{k+1}$ have the same sign,
for $k\neq 1,N$.
\end{itemize}

Then                                                                                

$$\|\Ss_E(\mathbf x)\|=\sum_{k=1}^N|(\Ss_E\mathbf x)_k|\le \sum_{k=1}^N|x_k|=\|\mathbf x\|,$$
and equality holds if and only if all $x_k$ have the same sign which is impossible for a non-zero
element of $\mathbf Z_E$.

\end{document}